\documentclass[prb,aps,twocolumn,showpacs,superscriptaddress]{revtex4-1}
\usepackage{graphicx} 
\usepackage{tabularx}
\usepackage{dcolumn} 
\usepackage{color}
\usepackage{bm}
\usepackage{amsmath}
\usepackage{amssymb}
\usepackage{multirow}

\newcommand{\nn}{\nonumber}

\begin{document} 

%\preprint{}

\title{
Nodal gap detection through polar angle-resolved density of states measurements in uniaxial superconductors
}



\author{Yasumasa Tsutsumi}  
\affiliation{
Department of Basic Science, University of Tokyo, Meguro, Tokyo 153-8902, Japan}
\affiliation{
Condensed Matter Theory Laboratory, RIKEN, Wako, Saitama 351-0198, Japan}
\author{Takuya Nomoto}  
\affiliation{
Department of Physics, Kyoto University, Kyoto 606-8502, Japan}
\author{Hiroaki Ikeda}  
\affiliation{
Department of Physics, Ritsumeikan University, Kusatsu, Shiga 525-8577, Japan}
\author{Kazushige Machida}  
\affiliation{
Department of Physics, Ritsumeikan University, Kusatsu, Shiga 525-8577, Japan}

%\author{Yasumasa Tsutsumi$^{1,2}$\thanks{tsutsumi@vortex.c.u-tokyo.ac.jp}, Takuya Nomoto$^3$, Hiroaki Ikeda$^4$, and Kazushige Machida$^4$} 
%\inst{
%$^1$Department of Basic Science, The University of Tokyo, Meguro, Tokyo 153-8902, Japan\\
%$^2$Condensed Matter Theory Laboratory, RIKEN, Wako, Saitama 351-0198, Japan\\
%$^3$Department of Physics, Kyoto University, Kyoto 606-8502, Japan\\
%$^4$Department of Physics, Ritsumeikan University, Kusatsu, Shiga 525-8577, Japan
%} 

%\author{Kazushige Machida} 
%\affiliation{
%Department of Physics, Ritsumeikan University, Kusatsu 525-8577, JAPAN}

\date{\today}

\begin{abstract}
We propose a spectroscopic method to identify the nodal gap structure in unconventional superconductors.
This method best suits for locating the horizontal line node and for pinpointing the isolated point nodes
by measuring polar angle ($\theta$) resolved zero energy density of states $N(\theta)$.
This is measured by specific heat or thermal conductivity at
low temperatures under a magnetic field.
We examine a variety of uniaxially symmetric nodal structure, including point and/or line nodes with linear and quadratic dispersions,
by solving Eilenberger equation in vortex states.
%We find (1) $N(\theta)$ exhibits a global maximum at $\theta_{max}$ started parallel to the anti-nodal direction which continuously shifts towards the
%nodal direction as field increases, thus (2) the oscillation pattern is reversed at low and high fields.
%(3) Just near the angles exactly pointing to the nodal direction a pair of the local minima may appear in $N(\theta)$.
%Various possible complications to hinder those three features are analyzed, such as the Fermi velocity anisotropy or multiband effects.
It is found that (A) the maxima of $N(\theta)$ continuously
shift from the anti-nodal to the nodal direction ($\theta_{\rm n}$)
as  a field increases accompanying the oscillation pattern reversal
at low and high fields.
Furthermore, (B) local minima emerge next to $\theta_{\rm n}$
on both sides except for the case of linear point node. These features are robust and detectable experimentally.
Experimental results of $N(\theta)$ performed on several superconductors, UPd$_2$Al$_3$, URu$_2$Si$_2$, Cu$_x$Bi$_2$Se$_3$, and UPt$_3$,
are examined and commented in light of the present theory.
\end{abstract}

%\pacs{pacs ???}
%74.	Superconductivity (for superconducting devices, see 85.25.-j)
%74.20.-z	Theories and models of superconducting state
%74.20.Pq	Electronic structure calculations (for methods of electronic structure calculations, see 71.15.-m)
%74.20.Rp	Pairing symmetries (other than s-wave)
%74.25.Bt	        Thermodynamic properties
%74.25.Ha	Magnetic properties including vortex structures and related phenomena (for vortices, magnetic bubbles, and magnetic domain structure, see 75.70.Kw)
%74.25.Jb	Electronic structure (photoemission, etc.)
%74.25.nj  	Nuclear magnetic resonance
%74.25.Op	Mixed states, critical fields, and surface sheaths
%74.25.Uv	Vortex phases (includes vortex lattices, vortex liquids, and vortex glasses)
%74.25.Wx	Vortex pinning (includes mechanisms and flux creep)
%74.62.En	Effects of disorder%
%74.70.Pq	Ruthenates
%61.05.fg	Neutron scattering (including small-angle scattering)

%\keywords{}


\maketitle

%%%%%%%%%%%%%%%%%%%%%%%%%%%%%%%

\section{Introduction}

The zero energy quasi-particles (QPs) bound in a vortex core play a crucial role in 
determining thermodynamics at low temperatures in 
various Fermion superfluids.~\cite{volovik1,volovik2} This is particular true for type II superconductors
both conventional and unconventional where external field generates vortices
accompanying low-lying Fermionic QPs in each vortex core.
The zero energy density of states (ZDOS) produced by those QPs can be probed by a variety of experimental techniques,
which sensitively reflects
the underlying gap structure, 
in particular, the nodal structure.
Their determination is
a foremost important task for identifying the unconventional pairing symmetry.~\cite{matsuda,sakakibara}

%There are a variety of the nodal gaps; for linear point nodes $\phi({\bm k})$$\propto$$\sqrt{k_x^2+k_y^2}$ and a linear line node $k_z$
%or quadratic points $k_x^2+k_y^2$, and line $k_z^2$
%and their combined hybrids in  unconventional superconductors~\cite{sigrist} where $\phi({\bm k})$ is the gap function.
%The position and direction of those nodes in reciprocal space are  decisive to identify the pairing symmetry of a material of interest.
%The line node runs not only vertically such as $d_{x^2-y^2}$ in high-$T_{\rm c}$ cuprates, but also runs horizontally like in chiral $d$-wave pairing 
%$(k_x+ik_y)k_z$.  There are still a lot of unconventional superconductors to be determined in its detailed nodal gap structure.

The nodes of the gap can be detected by low-field dependence of the ZDOS.
In contrast to full gap superconductors, in which the ZDOS is proportional to magnetic field $B$ owing to the density of vortices, the ZDOS in nodal superconductors depends on $\sqrt{B}$.~\cite{volovik1}
The field dependence mostly comes from the Doppler shifted QPs in the vicinity of the gap nodes where the gap is smaller than the Doppler shift energy $\delta E({\bm k})=m{\bm v}_{\rm F}({\bm k})\cdot{\bm v}_{\rm s}$.
Here, $m$ is the electron mass, ${\bm v}_{\rm F}({\bm k})$ is the Fermi velocity at ${\bm k}$ in the momentum space, and ${\bm v}_{\rm s}$ is the supercurrent velocity perpendicular to the field.
The Doppler shift energy for nodal QPs depends on the directions of the Fermi velocity near the nodes and the supercurrent, namely the relative orientation of the nodal position and field direction.
Therefore, the nodal positions are determined by modulation of the ZDOS under rotating magnetic field.
In low field region, the ZDOS shows minima when field is pointed to nodal directions,~\cite{vekhter} which is first demonstrated by using the semiclassical Doppler shift method.~\cite{kubert:1998}
The oscillation of the ZDOS for nodal gap structures is quantitatively clarified by the microscopically based quasiclassical theory with the Kramer--Pesch approximation (KPA)~\cite{nagai:2008b} which is valid under low fields.

Under high fields, the oscillation of the ZDOS should be reversed owing to the lowest upper critical field $B_{\rm c2}$ when field is pointed to nodal directions~\cite{miranovic1} [Fig.~1].
The crossover from the minima to the maxima of the ZDOS by magnetic field is demonstrated by the quasiclassical theory with the Brandt--Pesch--Tewordt (BPT) approximation~\cite{vorontsov:2006,vorontsov:2007b} which is valid under high fields near $B_{\rm c2}$.
The origin of the ZDOS inversion is QP scattering by magnetic field~\cite{vorontsov:2007b} which is lacked in the semiclassical Doppler shift method.

Thanks to the established theory, the azimuthal angle resolved density of states (DOS) measurements via  either specific heat~\cite{sakakibara} or thermal conductivity~\cite{matsuda} are available to determine position of vertical line nodes.
This method~\cite{vekhter,miranovic1,miranovic2,vorontsov:2006,vorontsov:2007b,vorontsov:2007c,hiragi} is quite effective to identify and distinguish the $d_{x^2-y^2}$ nodal structure from the
$d_{xy}$ symmetry, for example, by checking the sign change of the four fold oscillation pattern in the
temperature and field space.~\cite{an}

In contrast, there has been no established experimental method to detect horizontal line node positions or point nodes on the pole in momentum space.
The information concerning the uniaxially symmetric gap structures from the polar angle resolved ZDOS seems to be concealed by Fermi surface anisotropy in typical tetragonal or hexagonal superconductors.
Even for the non-uniaxially symmetric gap structures, we cannot readily distinguish between vertical line nodes and point nodes on the basal plane by the polar angle dependence of the ZDOS in low field region.~\cite{nagai:2008b}
Note that the determination of positions of point node for YNi$_2$B$_2$C was achieved by the ZDOS analysis considering a realistic Fermi surface obtained by a band calculation.~\cite{nagai:2007,nagai:2009}

This situation is contrasted with topological insulators or semimetals 
where angle resolved photoemission spectroscopy is powerful enough to directly map out the whole momentum space for Dirac or Weyl nodes.~\cite{nakatsuji}
In particular it seemed difficult to pinpoint the point node
position in spite of the recent intriguing proposals of Dirac and Weyl superconductors with linear or higher point nodes.~\cite{fu,fan,xu,haldane} 
This is only indirectly inferred from bulk thermodynamic measurements.
The difficulty of detection of point nodes is compounded by the fact that it often coexists with line nodes such as in gap function $\phi({\bm k})\propto(k_x+ik_y)k_z$
in which line nodes overwhelm point nodes in bulk thermodynamics.

In this paper, we aim to establish a spectroscopic method to detect point or/and horizontal line nodes by showing that the polar angle $(\theta)$ resolved ZDOS $N(\theta)$ contains valuable information on the nodal gap structure.
As schematically shown in Figs.~1(a) and 1(b) for typical horizontal line node and polar point node structures, respectively, on the Fermi sphere, $N(\theta=0^{\circ })$ and $N(\theta=90^{\circ })$ are generically crossed as a function of $B$.
%because  $N(B^{\parallel }\parallel$ node) is larger than $N(B^{\perp }\perp$ node) at lower $B$ while $B_{\rm c2}^{\parallel}<B_{\rm c2}^{\perp}$
%because  $B_{\rm c2}$ is proportional to the effective gap amplitude for that direction.
Thus the minimum and maximum positions of $N(\theta)$ under a fixed $B$ are reversed at the crossing field $B_{\rm CR}$.
Around this field the $N(\theta)$ pattern sensitively reflects the underlying nodal structure as we see below.

Indeed, the recent polar angle resolved specific heat measurements can detect a horizontal line node and point nodes in URu$_2$Si$_2$~\cite{kittaka} and a horizontal line node in UPd$_2$Al$_3$.~\cite{shimizu}
There have been only a few such systematic measurements in spite of the fact that nowadays those angle
resolved measurements have become a standard experimental technique.
It is our aim that we shed new light to the polar angle resolved ZDOS measurements and clarify their usefulness and limitations.

\begin{figure}
\begin{center}
\includegraphics[width=8cm]{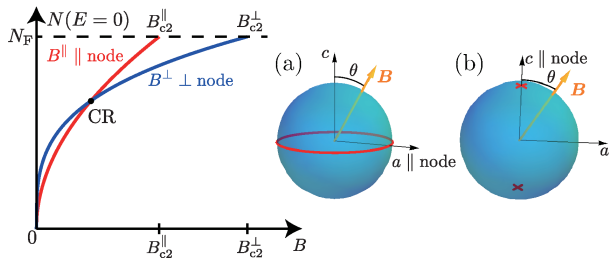}
\end{center}
\caption{\label{fig1}
Schematic figure of ZDOS as a function of $B$ for two directions.
CR indicates the crossing field of two ZDOS curves.
The oscillation pattern of $N(\theta)$ at a fixed field is reversed at around $B_{\rm CR}$.
(a) A line node at the equator and (b) point nodes at two poles in momentum space.
}
\end{figure}

\section{Quasiclassical Eilenberger theory and Kramer-Pesch approximation}

The polar angle resolved ZDOS $N(\theta)$ is microscopically derived by the quasiclassical Eilenberger theory.~\cite{eilenberger}
This framework is valid for superconductors with $k_{\rm F}\xi\gg1$ where $k_{\rm F}$ is the Fermi wave number and $\xi$ is
the coherence length, which is fulfilled by most superconductors of interest except perhaps for the high-$T_{\rm c}$ cuprates.
This framework is specially powerful for extracting the internal QP structure of the vortex lattice state in type II superconductors.
%Once the quasi-classical Green function is obtained by solving the Eilenberger equation numerically,
%one can gain various physical quantities~\cite{supplement,ichioka1,ichioka2,ichioka3}.  
%We are concerned here with ZDOS; 
%$N(E=0)=\left\langle N_{\bm k}(\bm{r},E=0)\right\rangle_{\bm{r},\bm{k}}$
%=$N_{\rm F}\!\left\langle{\rm Re}\left[g(\bm{k},\bm{r},\omega_n)|_{i\omega_n\rightarrow i0^+ }\right]\right\rangle_{\bm{r},\bm{k}}$,
%which is averaged over $\bm{r}$ and $\bm{k}$ space because this is directly measurable by specific heat and 
%thermal conductivity experiments at low temperatures.
%We focus on $N(E=0)$ and $N_{\bm k}(E=0)=\left\langle N_{\bm k}(\bm{r},E=0)\right\rangle_{\bm{r}}.$
%In particular the field-polar angle dependence of $N(\theta, E=0)$ or simply $N(\theta)$ is examined.
The QP state is obtained from the quasiclassical Green's functions $g\!\equiv\! g(\bm{k},\bm{r},\omega_n)$, $f\!\equiv\! f(\bm{k},\bm{r},\omega_n)$, and $\underline{f}\!\equiv\!\underline{f}(\bm{k},\bm{r},\omega_n)$ depending on the direction of the Fermi momentum $\bm{k}$, the center-of-mass coordinate $\bm{r}$ for the Cooper pair, and Matsubara frequency $\omega_n\!=\!(2n\!+\!1)\pi k_{\rm B}T$ with $n\!\in\!\mathbb{Z}$.
They are calculated in a unit cell of the triangle vortex lattice by solving the Eilenberger equation
\begin{equation}
\begin{split}
&\left\{\omega_n+\bm{v}_{\rm F}({\bm k})\cdot\left[\bm{\nabla }+i\bm{A}(\bm{r})\right]\right\}f=\Delta({\bm k},\bm{r})g,
\\
&\left\{\omega_n-\bm{v}_{\rm F}({\bm k})\cdot\left[\bm{\nabla }-i\bm{A}(\bm{r})\right]\right\}\underline{f}=\Delta^*({\bm k},\bm{r})g,
\end{split}\label{eq:Eilenberger}
\end{equation}
where $g\!=\!(1\!-\!f\underline{f})^{1/2}$, ${\rm Re}[g]\!>\!0$, and the order parameter consists of the gap value $\Delta({\bm r})$ and the gap function $\phi({\bm k})$ as $\Delta({\bm k},{\bm r})\!\equiv\!\Delta({\bm r})\phi({\bm k})$.
Instead of the rotation of magnetic field, we regard the Fermi velocity on the Fermi sphere as ${\bm v}_{\rm F}({\bm k})\!=\!v_{\rm F}\hat{R}^{-1}(\theta){\bm k}$, where $\hat{R}(\theta)$ is a rotation matrix through the polar angle $\theta$ about the $y$-axis.
Then, the magnetic field ${\bm B}\!=\!(0,0,B)$ is fixed to the $z$-direction and the vector potential is given by $\bm{A}(\bm{r})\!=\!\frac{1}{2}\bm{B}\!\times\!\bm{r}$ in the symmetric gauge.
Here, we assume the extreme type II superconductors with the large Ginzburg--Landau parameter.
The unit vectors of the triangle vortex lattice are given by $\bm{u}_1\!=\!(l,0,0)$ and $\bm{u}_2\!=\!(\frac{1}{2}l,\frac{\sqrt{3}}{2}l,0)$, where the lattice size is fixed by $\frac{\sqrt{3}}{2}l^2B\!=\!\phi_0$ with the flux quantum $\phi_0$.
%Throughout this paper, temperatures, energies, lengths, and magnetic fields are, respectively, measured in units of the transition temperature $T_{\rm c}$, $\pi k_{\rm B} T_{\rm c}$, $\xi_0\!=\!\hbar v_{\rm F}/2\pi k_{\rm B} T_{\rm c}$, and $B_0\!=\!\phi_0/2\pi\xi_0^2$.

The gap value is obtained by
\begin{align}
\Delta(\bm{r})=k_{\rm B}T\sum_{0<\omega_n\le\omega_{\rm c}}g_0N_{\rm F}\left\langle\phi^*({\bm k})(f+\underline{f}^*)\right\rangle_{\bm{k}}\label{eq:gap}
\end{align}
where $\langle\cdots\rangle_{\bm{k}}$ indicates the Fermi surface average.
The coupling constant $g_0$ and the DOS in the normal state $N_{\rm F}$ have the relation $(g_0N_{\rm F})^{-1}\!=\!\ln (T/T_{\rm c})\!+\!k_{\rm B}T\sum_{|\omega_n|\le\omega_{\rm c}}\omega_n^{-1}$, where $T_{\rm c}$ is the transition temperature.
Here, we use the energy cutoff $\omega_{\rm c}\!=\!20k_{\rm B}T_{\rm c}$.
We self-consistently calculate the quasiclassical Green's functions and the gap value $\Delta(\bm{r})$ under a given unit cell of the triangle vortex lattice with the periodic boundary condition including the phase factor due to the magnetic field.~\cite{ichioka1,ichioka2,ichioka3}
Although the vortex lattice configuration is distorted from an equilateral triangle when magnetic field is tilted from the $c$-axis for uniaxially symmetric gap functions, we can safely neglect the deformation because it is small unless the high field or high temperature region even for a two-fold symmetric gap function.~\cite{tsutsumiUPt3}

When we calculate the QP state, we solve Eq.~\eqref{eq:Eilenberger} under the self-consistent gap value $\Delta({\bm r})$ with $i\omega_n\!\rightarrow\! E\!+\!i\eta$.
The momentum-resolved DOS is given by
\begin{align}
N_{\bm k}(E)\!=\!N_{\rm F}\!\left\langle{\rm Re}\left[g(\bm{k},\bm{r},\omega_n)|_{i\omega_n\!\rightarrow\! E\!+\!i\eta }\right]\right\rangle_{\bm{r}}\!,
\end{align}
where $\langle\cdots\rangle_{\bm r}$ indicates the spatial average in the unit cell of the vortex lattice.
Here, we typically use the smearing factor $\eta\!=\!0.003\pi k_{\rm B}T_{\rm c}$.
The DOS is obtained by the Fermi surface average of the momentum-resolved DOS as $N(E)\!=\!\langle N_{\bm k}(E)\rangle_{\bm{k}}$.
We focus on the ZDOS $N(E=0)$ and momentum-resolved ZDOS $N_{\bm k}(E=0)$ when field is tilted from the $c$-axis with the polar angle $\theta$.
The ZDOS and momentum-resolved ZDOS are abbreviated to $N(\theta)$ and $N_{\bm k}(\theta)$, respectively.

We should confirm the robustness of features in the polar angle resolved ZDOS $N(\theta)$ against Fermi surface anisotropy because our aim is to single out nodal information from $N(\theta)$.
However, since it is too heavy to solve the Eilenberger equation self-consistently for several anisotropic Fermi surfaces, we use the Kramer--Pesch approximation (KPA)~\cite{Nagai2006,nagai:2008b,nagai:2009,Nagai2011} which is valid under low fields.
Within the KPA, we can obtain a reasonable solution of Eq.~\eqref{eq:Eilenberger} without a heavy 
numerical calculation. In the KPA with the Riccati formalism, a one-vortex solution of Eq.~\eqref{eq:Eilenberger} is given by~\cite{Nagai2011}
\begin{align}
\frac{N(\bm{r},E=0)}{N_F}=\left\langle\frac{v_{\perp}(\bm{k})e^{-u(s)}}{\pi C(y,\bm{k})}\frac{\eta}{E^2(y,\bm{k})+\eta^2}\right\rangle_{\bm{k}},
\end{align}
where $\bm{v}_{\perp}(\bm{k})$ is a projection of $\bm{v}_{\rm F}(\bm{k})$ into the plane perpendicular to the field and $(s,y)$ is a coordinate of the plane in $\bm{v}_{\perp}(\bm{k})$. When we parameterize $\Delta(\bm{r})=f(s,y)e^{i\phi_r}$ with an angle $\phi_r$ around a vortex, $u(s), C(y,\bm{k})$, and $E(y,\bm{k})$ can be expressed by $f(s,y)$ and given analytically for an appropriate $f(s,y)$. Note that in the one-vortex approximation, we cannot take into account vortex lattice formation and the magnetic field effect only appears as an integral radius of $\!\langle N(\bm{r},E=0)\rangle_{\bm{r}}$.
%In the KPA calculations, we deal with the effect of the Fermi velocity anisotropy $\Gamma$ defined below.
%The change of the coherence length $\xi_{0\perp }$ in the plane perpendicular to the field is estimated by the second derivative terms of the Ginzburg--Landau free-energy. 
%$\Delta(\bm{k})|_{\rm max}=1,B_{c2}^{\parallel c} =1$, and $\eta = 0.05$.
We confirm that KPA results coincide qualitatively with those by the Eilenberger full solution.

The Fermi surface anisotropy of a uniaxial superconductor is introduced by the anisotropy parameter $\Gamma_0$ defined by
\begin{align}
\Gamma_0\equiv\sqrt{\frac{\langle v_{ab}({\bm k})^2\rangle_{\bm k}}{\langle v_c({\bm k})^2\rangle_{\bm k}}},
\end{align}
where $v_{ab}({\bm k})$ and $v_c({\bm k})$ are the Fermi velocity along the $ab$-plane and the $c$-axis, respectively.
Within the effective mass model, the polar angle dependence of the Fermi velocity is described by
\begin{align}
\frac{v_{\rm F}(\theta)}{v_{\rm F}(0^{\circ })}=\sqrt{\Gamma_0^2\sin^2\theta+\cos^2\theta }.
\end{align}

For the isotropic superconducting gap, the Fermi surface anisotropy leads to the anisotropy of the upper critical field as
\begin{align}
\frac{B_{\rm c2}(\theta)}{B_{\rm c2}(90^{\circ })}=\frac{1}{\sqrt{\Gamma_0^2\cos^2\theta+\sin^2\theta }}.
\end{align}
Since the ZDOS is linear in $B$ at low fields,
each $N(\theta)$ points to its own $B_{\rm c2}(\theta)$ by one to one correspondence.
At a given $B$, $N(\theta)$ behaves inversely proportional to $B_{\rm c2}(\theta)$
except for higher fields where non-linear behavior becomes apparent,~\cite{nakai} namely
\begin{align}
\frac{N(\theta)}{N(90^{\circ })}=\sqrt{\Gamma_0^2\sin^2\theta+\cos^2\theta }.
\end{align}
It is clear that $N(\theta)/N(90^{\circ })$ is independent of $B$ at least in low fields
and monotonic function of $\theta$, which is not universal for nodal gap structures.

For uniaxially symmetric nodal gap structures, the upper critical field depends on the general anisotropy parameter $\Gamma$ as
\begin{align}
\frac{B_{\rm c2}(\theta)}{B_{\rm c2}(90^{\circ })}=\frac{1}{\sqrt{\Gamma^2\cos^2\theta+\sin^2\theta }},
\end{align}
where $\Gamma$ is defined by
\begin{align}
\Gamma\equiv\sqrt{\frac{\langle[\phi({\bm k})v_{ab}({\bm k})]^2\rangle_{\bm k}}{\langle[\phi({\bm k})v_c({\bm k})]^2\rangle_{\bm k}}}.
\end{align}
In the KPA calculations, the change of the coherence length in the plane perpendicular to the field is estimated by the relation:
\begin{align}
\Gamma=\frac{B_{\rm c2}^{ab}}{B_{\rm c2}^c}=\frac{\xi_{ab}}{\xi_c},
\end{align}
where $B_{\rm c2}^{ab(c)}$ and $\xi_{ab(c)}$ are the upper critical field and the coherence length along the $ab$-plane ($c$-axis), respectively.

\section{Polar angle resolved ZDOS}

\subsection{Self-consistent results by the Eilenberger theory}

\begin{figure}
\begin{center}
\includegraphics[width=8.5cm]{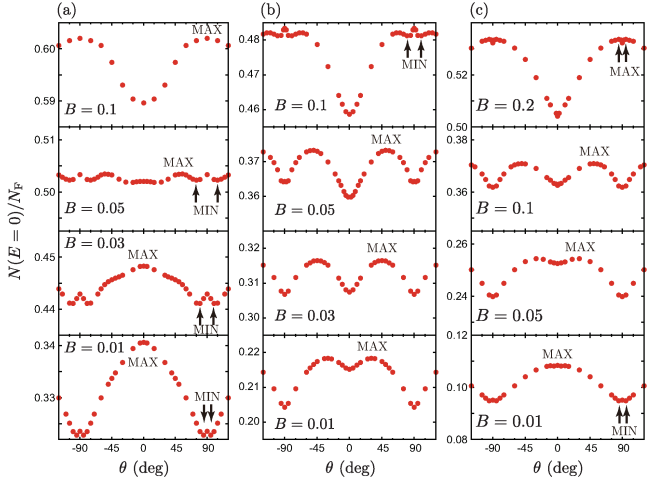}
\end{center}
\caption{\label{fig2}
Field evolutions of polar angle resolved ZDOS normalized by the normal DOS $N_{\rm F}$ 
for  (a) $\phi({\bm k})\propto k_z^2$,  (b) $k_z^2(k_x^2+k_y^2)$, and (c) $k_z(k_x+ik_y)$.
The magnetic field is scaled by the Eilenberger unit $B_0=\phi_0/2\pi\xi^2$.
}
\end{figure}

First, we show the polar angle resolved ZDOS obtained from the self-consistent calculation using the Eilenberger theory.
Figures 2(a), 2(b), and 2(c) show the ZDOS for the gap functions $\phi({\bm k})\propto k_z^2$, $k_z^2(k_x^2+k_y^2)$, and $k_z(k_x+ik_y)$, respectively, at $T=0.2T_{\rm c}$. 
We first focus on Fig.~2(a) for  $\phi({\bm k})\propto k_z^2$ with a quadratic line node case.
Starting with $N(0^{\circ})>N(90^{\circ})$ at low fields, $N(\theta)$ evolves and changes its oscillation patterns upon increasing $B$, which is scaled by the Eilenberger unit $B_0=\phi_0/2\pi\xi^2$.
The oscillation pattern is reversed, namely $N(0^{\circ})<N(90^{\circ})$ at higher fields through $N(0^{\circ})=N(90^{\circ})$, corresponding 
to the crossing field (CR) in Fig.~1.
%This oscillation pattern reversal is generic and common for all three nodal structures.
During this field evolution the global maximum denoted by MAX in Fig.~2(a) continuously moves towards $\theta=\pm 90^{\circ}$ from $\theta=0^{\circ}$.
It is understood that the oscillation reversal is driven by moving the global maxima, or the ``MAX structure". 
The moving MAX structure in the oscillation reversal fields is more noticeable for the hybrid nodal gap structures:
$\phi({\bm k})\propto k_z^2(k_x^2+k_y^2)$ with 
quadratic point nodes and a quadratic line node in Fig.~2(b) and $\phi({\bm k})\propto k_z(k_x+ik_y)$ with linear point and line nodes
in Fig.~2(c). 

In addition to the MAX structure, we can see the local minima
denoted by MIN just near $\theta=90^{\circ}$ in Fig.~2(a). This ``MIN structure" just
near the line nodal position is commonly seen in Figs.~2(b) and 2(c) albeit the visibility and field region depend on the nodal structure.
Generally the MIN structure in the quadratic line node case is easier to see than that in the linear line node case as found by the comparison of Figs.~2(a) and 2(c).
This is simply because the amount of the excited QPs is larger in a quadratic line node than that in a linear line node.~\cite{phasespace}
%And the line node is easier to see than the point node.
%Those are simply because the phase space volumes~\cite{phasespace} of the nodal QPs are different for those cases.

Thus far, we have found the two common features in $N(\theta)$:
(A) the moving global maxima, MAX structure, and (B) the local minima just near the node, MIN
structure.
Those findings are non-trivial to understand from simple Doppler shift picture.~\cite{volovik1,vekhter}
In what follows we discuss the physical origins in terms of the QP picture to check their generality and limitations.
For the qualitative comprehension, the ZDOS estimated from the order parameter in the vortex state described only by the lowest Landau level,~\cite{vorontsov:2007b}
\begin{align}
N(E=0)\approx&\left\langle\left[1+2\left(\frac{2\Lambda\Delta({\bm k})}{v_{\perp }({\bm k})}\right)^2\right]^{-\frac{1}{2}}\right\rangle_{\bm k}\nn\\
=&\left\langle\left[1+\frac{1}{4z^2}\left(\frac{\phi({\bm k})}{v_{\perp }({\bm k})/v_{\rm F}}\right)^2\right]^{-\frac{1}{2}}\right\rangle_{\bm k},
\label{eq:LLL}
\end{align}
is helpful, where $z\equiv v_{\rm F}/(4\sqrt{2}\Lambda\Delta)\sim\xi/\Lambda\sim\sqrt{B/B_{\rm c2}}$ with the distance $\Lambda$ between vortices.

%For the gap function $\phi({\bm k})\propto k_z^2$ with a quadratic line node [Fig.~1(a)], there are local minima of the ZDOS in the vicinity of $\theta=\pm 90^{\circ }$ below $B=0.1$.
%Also for the gap function $\phi({\bm k})\propto k_z^2(k_x^2+k_y^2)$ which has quadratic point nodes in addition to a quadratic line node [Fig.~1(b)], local minima of the ZDOS are present near $\theta=90^{\circ }$ at $B=0.1$.
%In the case of the gap function with a linear line node and linear point nodes, $\phi({\bm k})\propto k_z(k_x+ik_y)$ [Fig.~1(c)], there are slight local minima of the ZDOS near the line node at the low field $B=0.01$.
%Thus, the local minima of the polar angle resolved ZDOS accompanies the horizontal line node regardless of the linear or quadratic node.

\begin{figure}
\begin{center}
\includegraphics[width=7cm]{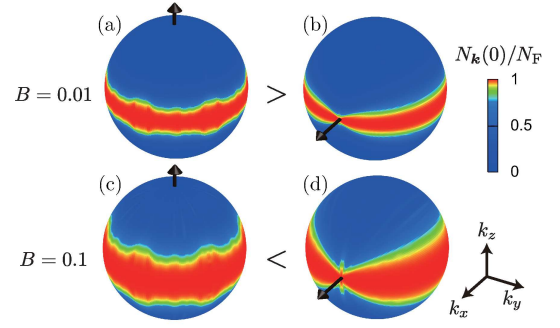}
\end{center}
\caption{\label{fig3}
Momentum-resolved ZDOS $N_{\bm k}(E=0)$ on the Fermi sphere for $\phi({\bm k})\propto k_z^2$.
The arrows indicate the field direction.
(a) $B$=0.01, $\theta$=0$^{\circ}$,
(b) $B$=0.01, $\theta$=90$^{\circ}$,
(c) $B$=0.1, $\theta$=0$^{\circ}$, and
(d) $B$=0.1, $\theta$=90$^{\circ}$
}
\end{figure}

The oscillation pattern reversal has been clarified by the quasiclassical theory with the BPT approximation, which is due to the QP scattering by magnetic field.~\cite{vorontsov:2007b}
It is confirmed from the momentum-resolved ZDOS $N_{\bm k}(E=0)$ for $\phi({\bm k})\propto k_z^2$, as shown in Fig.~3.
At low fields, the spectral weight from the line node 
for $B\parallel c$ in Fig.~3(a) overwhelms that for $B\perp c$ in Fig.~3(b), resulting in $N(0^{\circ})>N(90^{\circ})$. 
%Note that the total ZDOS is given by integrating those weights over the Fermi surface.
The zero energy QPs appear in the narrow region around the equator.
%The Doppler shift restriction $\Delta\phi({\bm k})<{\bm v}_{\rm s}\cdot {\bm p}_{\rm F}$, 
%where $\Delta$ is the superconducting gap, ${\bm v}_{\rm s}$ $(\perp {\bm B})$ is the supercurrent velocity induced by field, and ${\bm p}_{\rm F}$ is the Fermi momentum, allows to excite QPs in the red belt.
For $B\perp c$ in Fig.~3(b), the momentum space region around the field direction 
is prohibited from the QP excitations because the Doppler shift energy $\delta E({\bm k})=m{\bm v}_{\rm F}({\bm k})\cdot{\bm v}_{\rm s}$ is small.
This is perfectly matched with the simple Doppler shift picture.~\cite{volovik1,vekhter}
On the other hand, the zero energy QPs are excited beyond the nodal region at high fields.
At high fields, since $z$ in Eq.~\eqref{eq:LLL} becomes order unity, finite $\Delta({\bm k})$ region contributes to the ZDOS except for the vicinity of the field direction with small $v_{\perp }({\bm k})$.
Therefore, the excited region is more enlarged for $B\perp c$ [Fig.~3(d)] than $B\parallel c$ [Fig.~3(c)].
%the Doppler shift restriction ${\bm v}_{\rm s}\cdot {\bm v}_{\rm F}$ is now to prohibit the QP excitations for $B\parallel c$
%while for $B\perp c$ it allows to cover the larger momentum region as seen from Fig.~3(d).
%On the other hand, the higher field is not simple.
%Since the allowable zero energy QPs are extended beyond the nodal region at high fields,
%the Doppler shift restriction ${\bm v}_{\rm s}\cdot {\bm v}_{\rm F}$ is now to prohibit the QP excitations for $B\parallel c$
%while for $B\perp c$ it allows to cover the larger momentum region as seen from Fig.~3(d).
This implies not only  $N(0^{\circ})<N(90^{\circ})$, but also  $B^a_{\rm c2}<B^c_{\rm c2}$.
This explanation holds for other nodal structures, including the point node case [see Fig.~6].

The moving global maxima, MAX structure (A) accompanies the oscillation pattern reversal.
For the hybrid nodal gap with point and line nodes in Figs.~2(b) and 2(c), the MAX structure is clearly seen because the QP scattering by magnetic field is enhanced especially around the point nodes when field is tilted from the $c$-axis.~\cite{kittaka}
Even for $\phi({\bm k})\propto k_z^2$ with a line node in Fig.~2(a), the slight MAX structure is seen near the crossing field.
The point node case also shows the MAX structure as mentioned below [see Fig.~6].
Thus, the MAX structure (A) is universal; however, that has not been reported by the azimuthal angle resolved experiment.
For instance, the azimuthal angle resolved ZDOS for $d_{x^2-y^2}$ nodal structure on the cylindrical Fermi surface is estimated from Eq.~\eqref{eq:LLL}, in which the crossing field corresponds to $z\approx 0.63$.~\cite{vorontsov:2007b}
The estimated ZDOS shows the moving global maxima in field region $0.60<z<0.67$ although the ratio $N(\phi_{\rm min})/N(\phi_{\rm max})\lesssim 0.993$, where $\phi_{\rm max(min)}$ is the azimuthal angle gives the maximum (minimum) ZDOS.
The oscillation of the ZDOS including the MAX structure less than $1\%$ is too small to detect the MAX structure.
In contrast, the oscillations of the polar angle resolved ZDOS for $\phi({\bm k})\propto k_z^2(k_x^2+k_y^2)$ at $B=0.05$ [Fig.~2(b)] and for $\phi({\bm k})\propto k_z(k_x+ik_y)$ at $B=0.1$ [Fig.~2(c)] are about $3.6\%$ and $2.4\%$, respectively.

\begin{figure}
\begin{center}
\includegraphics[width=6cm]{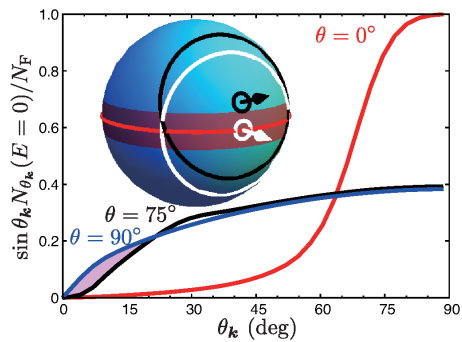}
\end{center}
\caption{\label{fig4}
Weighted ZDOS $\sin \theta_{\bm k} N_{\theta_{\bm k}}(E=0)$ for selected field angles for $\phi({\bm k})\propto k_z^2$ at $B=0.03$.
Inset shows schematic configuration of the line node on the Fermi sphere and two field directions $75^{\circ }$ (black arrow) 
and $90^{\circ }$ (white arrow) with two integration paths over $\phi_{{\bm k}}$ for $\theta_{{\bm k}}=5^{\circ }$ (small circles) and $45^{\circ }$ (large circles).
}
\end{figure}

The physical understanding of the appearance of the minima near the nodal position, MIN structure (B), is explained in Fig.~4 where 
the weighted ZDOS $\sin\theta_{{\bm k}}N_{\theta_{\bm k}}(E=0)$ is shown for the selected $\theta$ values of $\phi({\bm k})\propto k_z^2$ at $B=0.03$.
As shown in Fig.~2(a), the MIN structure is clearly seen at $\theta\approx 75^{\circ }$ in this field.
Here, $N_{\theta_{\bm k}}(E=0)\equiv\frac{1}{2\pi }\int_0^{2\pi }d\phi_{\bm k}N_{\bm k}(E=0)$ and $\theta_{\bm k}$ is measured from the field direction.
Note that the area under each curve yields $N(\theta)/2$, which corresponds to the maximum in $N(0^{\circ})$ in this field.
It is seen from Fig.~4 that the largest contribution comes near $\theta_{{\bm k}}=90^{\circ}$ for $\theta=0^{\circ}$ case, namely around the line node.

The occurrence of  the MIN structure is understood by comparing two curves for $\theta=75^{\circ}$
and $\theta=90^{\circ}$ in Fig.~4.  As emphasized by the shaded region, the main area difference which ultimately results in the local minimum at  $\theta\approx 75^{\circ}$
comes from those small $\theta_{\bm k}$'s.
The deficiency of the area for $\theta=75^{\circ}$ compared with that for $\theta=90^{\circ}$ is easily found by looking at the inset
where two integration paths over $\phi_{\bm k}$ at $\theta_{\bm k}$=5$^{\circ}$ (small circles) and 45$^{\circ}$ (large circles) 
for two field directions, $\theta=90^{\circ}$ and $75^{\circ}$ indicated by arrows.
The white small circle for $\theta=90^{\circ}$ is situated inside the belt around the line node where the excitable QPs are abundant
while the black small circle for $\theta=75^{\circ}$ just misses to hit this region.
In contrast, the two larger white and black circles for higher  $\theta_{\bm k}$ give similar contributions to ZDOS.
Note that the region with the excited QPs, schematically depicted as a belt around the line node, is enlarged by increasing the magnetic field.
Then, the polar angles giving the local minima shift apart from the nodal direction by increasing $B$, as shown in Fig.~2(a).
% For $\theta=60^{\circ }$, the ZDOS is not gained by QPs with small $\theta_{{\bm k}}$ (the smaller black circle). 
%However, the ZDOS has a hump at $\theta_{\bm k}=45^{\circ }$ because the integration pass begins hitting the belt with long sections (the larger black circle).

We have explained the occurrence of the MIN structure just near the nodal direction for the
quadratic line node case in Fig.~2(a).  As anticipated by the above physical reasoning,
the MIN structure can be seen in other nodal structures. Indeed we can see it in Figs.~2(b) and 2(c)
for the hybrid nodal cases with point and line nodes and also isolated linear line node case in Fig.~5. 
%Also as shown in supplement part,
%the MIN structure is present for the quadratic point node (see Fig.~S1).
The MIN structure is also present for the quadratic point node case [see Fig.~6].
Moreover, the MIN structure in thermal conductivity was reported by the model calculation in multi-band nodal superconductors.~\cite{mishra:2011}
Thus it is quite universal and robust against possible complications as discussed next.

\subsection{Confirmation with the Kramer--Pesch approximation}

Here, we confirm the robustness of our assertion consisting of (A) the MAX and (B) the MIN structures
associated with the nodal gap.
In order to check that, we examine various cases by the quasiclassical theory with the KPA, such as the Fermi velocity anisotropy or topology of the
nodal gap structure, including the linear or quadratic point nodes.
We also examine off-symmetric nodal structures.
%Here, we describe the effect of the Fermi velocity anisotropy and some of other situations are
%described in Supplemental Material~\cite{supplement}.

\subsubsection{Fermi surface anisotropy}

%So far we have only considered the situation that the Fermi velocity is isotropic.
The anisotropy of the Fermi velocity is taken into account by modifying the 
effective masses for two perpendicular directions, $ab$- and $c$-axes,
resulting in the deformed Fermi surfaces depending on the anisotropy parameter $\Gamma_0=v_{ab}/v_c$.
We examine two Fermi surface cases, a cigar type elongated along the $c$-axis
where the anisotropy parameter becomes larger than unity ($\Gamma_0=1.2$)
and a spheroid type with the anisotropy parameter smaller than unity ($\Gamma_0=0.71$).
Those Fermi surface models are conveniently evaluated by the KPA,~\cite{Nagai2006,Nagai2011}
giving the results shown in Fig.~5 for the horizontal linear line node $\phi({\bm k})\propto k_z$.
The MIN structure in the Fermi sphere case [Fig.~5(b)] is hard to see compared to that for $\phi({\bm k})\propto k_z^2$ with a quadratic line node in Fig.~2(a) because the QP excited area around the node is narrow for the linear line node.~\cite{phasespace}
Then, the polar angles giving the local minima are very close to $\theta=90^{\circ }$, which shift apart in high fields.
Even for the linear line node case, however, the MIN structure is clearly seen in cigar type Fermi surfaces, as shown in Fig.~5(a).
The anisotropy parameter $\Gamma_0=v_{ab}/v_c>1$ provides the enlarged QP excited area around the line node, which is schematically shown in the inset of Fig.~4.
Then, the MIN structure is clear for cigar type Fermi surfaces; in contrast, that is unclear for spheroid type Fermi surfaces as shown in Fig.~5(c).
Also the MAX structure is easily observed in cigar type Fermi surfaces owing to the extension of the field range where the oscillation pattern is reversed.
Although the visibility of the MAX and MIN structures depends on the Fermi velocity anisotropy, all our recognized features are present and robust against the anisotropy.
%Whether the MIN structure is easily or indistinctly seen, two features,
%(B) the global maxima, MAX, move to $\theta=90^{\circ}$ with increasing $B$ and
%(C) the local minima, MIN, appear just near the line node position at $\theta=90^{\circ}$,
%are present and robust against the Fermi velocity anisotropy.
%The Fermi sphere case in middle is qualitatively same as in the previous results by
%full Eilenberger solution. It is clearly seen from Fig.~5 that two characteristics
%(B) the global maxima MAX move to $\theta=90^{\circ}$ with increasing $B$,
%(C) the local minima MIN appear just near the line node position at $\theta=90^{\circ}$
%are present and robust against the Fermi velocity anisotropy changes.

We should notice a situation carefully when we try to detect the nodal gap structure
experimentally. As shown in Fig. 1 schematically, the two $N(\theta)$ curves generically 
cross at a certain field $B_{\rm CR}$ for the isotropic Fermi velocity.
For the horizontal line node (polar point nodes) case, as the Fermi velocity anisotropy $\Gamma_0$ increases (decreases)
by modifying the Fermi sphere to the cigar (spheroid) type, the crossing field CR shifts to higher field region
and at a certain $\Gamma_0$ it disappears. (The opposite case is not problematic since it preserves the CR, shifting it to lower $B$; thus, in principle
we can have a CR always.) In this situation it becomes hard to see the MAX (A) and MIN (B)
structures because those could be observed in the field region where two
values of $N(0^{\circ})$ and $N(90^{\circ})$ are comparable.
%Thus interesting and useful information
%for the nodal structure is limited to lower field below which $N(\theta)$ is dominated by the
%Fermi velocity anisotropy.
%This is the case in URu$_2$Si$_2$~\cite{kittaka} where the useful nodal information on the MAX structure (B)
%is indeed obtained in low field region. In contrast in UPd$_2$Al$_3$~\cite{shimizu} the full range of field region provides
%useful nodal gap information, including the oscillation reversal (A) and the MAX structure (B) associated with a horizontal line node.
%This is because in UPd$_2$Al$_3$ the upper critical fields $B_{\rm c2}^{ab}$ and $B_{\rm c2}^{c}$ are comparable, implying that
%the Fermi velocity anisotropy is small.

\begin{figure}
\begin{center}
\includegraphics[width=8.5cm]{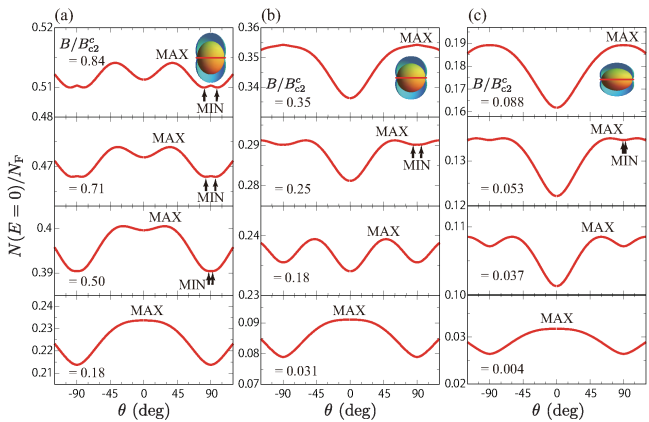}
\end{center}
\caption{\label{fig5}
Field evolutions of $N(\theta)$ within the KPA for three Fermi surfaces, cigar (a), sphere (b), and spheroid (c)
corresponding to the anisotropy parameters $\Gamma_0=1.2$, $1$, and $0.71$,
respectively. The linear line node is on the equator.
}
\end{figure}

\subsubsection{Linear and quadratic point nodes}

Here we show the comparison of two cases with linear and quadratic point nodes on the Fermi sphere, namely 
$\phi({\bm k})\propto \sqrt{k_x^2+k_y^2}$ and  $\phi({\bm k})\propto k_x^2+k_y^2$ 
in Figs.~6(a) and 6(b), respectively. For both cases the MAX structure is present. 
It is seen that the minima just near $\theta=0^{\circ}$ appear
 for the quadratic case in certain field region while they are absent for the linear case.
This enables us to distinguish those two point nodal structures
owing to the different amount of the nodal QPs.
%This example in addition to the linear and quadratic lines demonstrated in the main text
%facilities to establish the local or global minimum structure appeared just near the nodal 
%position. This MIN structure can be observed even in the hybrid nodal cases
%with the point and  line nodes as seen in Figs.~2(b) and 2(c).

In order to observe the distinctive MIN structure (B) which is able to pinpoint the nodal position in momentum space,
it is advantageous for the node to have larger amount of the QPs.~\cite{phasespace}
Then, it is possible to see the MIN structure for the quadratic point node as demonstrated in Fig.~6(b); however, the MIN structure is absent around the linear point node as shown in Fig.~6(a).
This remark is specially important because there is no alternative spectroscopic method
to pinpoint the point node in momentum space, in view of the recent various interesting predictions~\cite{fu,fan,xu,haldane} as for Dirac node
and Weyl node with the linear or higher order dispersion in topological superconductors.
Moreover, the order of the dispersion can be known from presence or absence of the MIN structure.
%Note that the ``failed'' linear point node with a small finite gap behaves like a quadratic point node.

In contrast, the hybrid gap consisting of the point and line nodes in Figs.~7(a) and 7(b) corresponding to Figs.~2(c) and 2(b), respectively, does not show the MIN structure associated with the point nodes because the coexisting line node overwhelms point nodes.
Meanwhile, the MAX structure is easily seen in the hybrid gap owing to the extension of the field range where the oscillation pattern is reversed.
The extension of the field range is similar to the solely line node case on cigar type Fermi surfaces [Fig.~5(a)]; however, the visibility of the MIN structure associated with the line node is not improved in the hybrid gap which is contrasted to the Fermi surface anisotropy.

\begin{figure}
\begin{center}
\includegraphics[width=8cm]{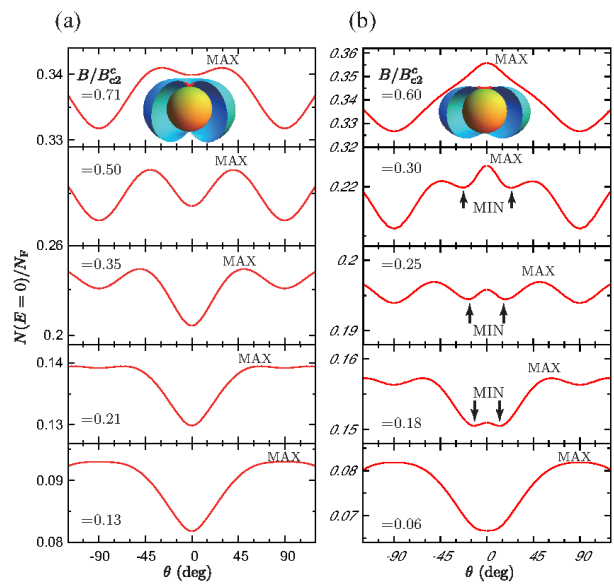}
\end{center}
\caption{\label{fig6}
Field evolutions of $N(\theta)$ within the KPA for the linear point nodes $\phi({\bm k})\propto \sqrt{k_x^2+k_y^2}$ (a) 
and quadratic point nodes $\phi({\bm k})\propto k_x^2+k_y^2$ (b).
}
\end{figure}

\begin{figure}
\begin{center}
\includegraphics[width=8cm]{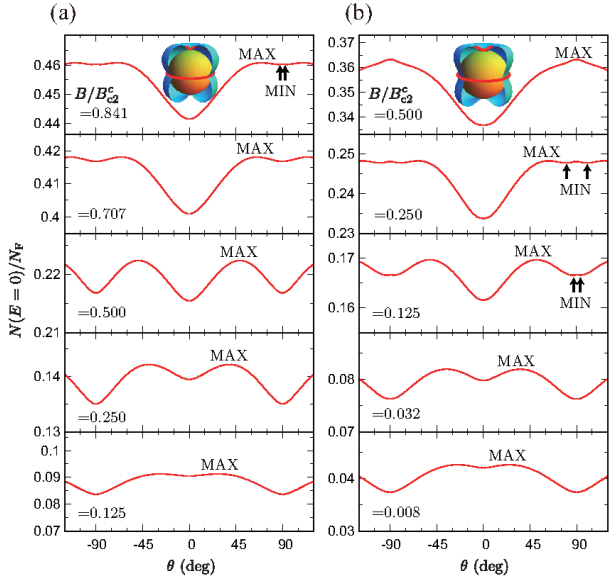}
\end{center}
\caption{\label{fig8}
Field evolutions of $N(\theta)$ within the KPA for the linear point and line nodes $\phi({\bm k})\propto k_z\sqrt{k_x^2+k_y^2}$ (a) 
and quadratic point and line nodes $\phi({\bm k})\propto k_z^2(k_x^2+k_y^2)$ (b).
}
\end{figure}

\subsubsection{Off-symmetric nodes}

We examine the MIN structure for the linear line nodes which is situated at off-symmetric positions in momentum space, namely away from the equator on the Fermi sphere.
As seen from Fig.~8 for the gap function $\phi({\bm k})\propto (5k_z^2-1)$,
the MIN structure is clearly demonstrated.
However, a pair of local minima associated with one linear line node around $\theta_{\rm n}$ is not seen, where $|\theta_{\rm n}|=\cos^{-1}(1/\sqrt{5})\approx 63^{\circ }$ is the polar angle for the line node.
A local minimum on the equator side of $\theta_{\rm n}$ is left but that on the pole side is concealed by the MAX structure.
Finally, the remaining local minimum is wiped off by the moving maximum toward $\theta=90^{\circ }$ in high fields.
%Here, however, a pair of local minima associated with one linear line node are merged into one
%local or global minimum because the two line nodes at $k_z=\pm 1/\sqrt5$ on the Fermi sphere are
%too strongly influenced each other to separately give rise to the four split local minima, leaving only two minima.

Thus it is understood that in order to clearly see a pair of local minima,
the position of the MAX structure should be situated apart from the concerned node.
%the nodal structure should be isolated from other nodes in momentum space.
This is one reason why, in the hybrid gap consisting of the quadratic point and line nodes shown in
Fig.~2(b), the MIN structure associated with the quadratic point node near $\theta=0^{\circ}$
is difficult to be resolved.

\begin{figure}
\begin{center}
\includegraphics[width=5cm]{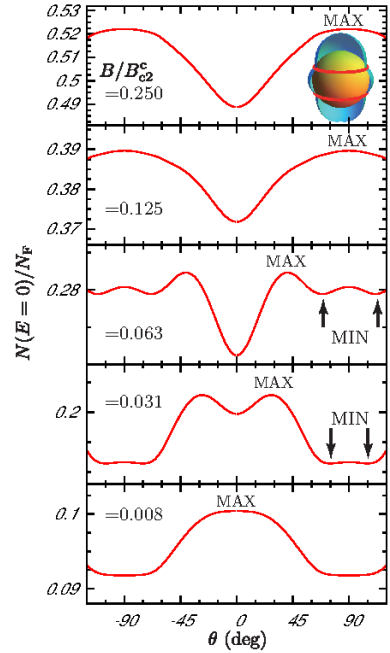}
\end{center}
\caption{\label{fig7}
Field evolution of $N(\theta)$ within the KPA for the off-symmetric line nodes $\phi({\bm k})\propto (5k_z^2-1)$. 
}
\end{figure}

\section{Considerations on experiments}

We discuss probable nodal structures for UPd$_2$Al$_3$,~\cite{shimizu} URu$_2$Si$_2$,~\cite{kittaka} Cu$_x$Bi$_2$Se$_3$,~\cite{yonezawa} and UPt$_3$,~\cite{machida} in which the polar angle resolved specific heat or thermal conductivity measurement was performed recently, based on our finding features:
(A) the moving MAX structure and (B) the local MIN structure just near the node.
Note that the amplitude of the ZDOS oscillation with the local MIN structure at $\theta_{\rm MIN}$ is estimated as $N(\theta_{\rm MIN})/N(\theta=0^{\circ })=0.984$ and $N(\theta_{\rm MIN})/N(\theta=90^{\circ })=0.996$ for the quadratic line nodal structure in Fig.~2(a) at $B/B_0=0.03$.
For the quadratic point nodal structure in Fig.~6(b) calculated with the KPA, $N(\theta_{\rm MIN})/N(\theta=0^{\circ })=0.994$ and $N(\theta_{\rm MIN})/N(\theta_{\rm MAX})=0.988$ at $B/B_{\rm c2}^c=0.25$, where $\theta_{\rm MAX}$ is the polar angle giving the maximum of the ZDOS.
Thus, although the MIN structure is marginal to the detection in experiments, the flat-bottomed ZDOS around the node will be observed at least.
The smaller MIN structure near a linear line node shown in Fig.~5 can be hardly detected except on a cigar type Fermi surface. 

%If the existence of the nodal gap is imposed by symmetry, not accidentally
%the foregoing arguments are applied to multiband superconductors
%because the multi-sheeted Fermi surfaces develop the identical nodal structure
%and we can map those multi-sheets into an effective single Fermi surface discussed here.

\subsection{UPd$_2$Al$_3$}

The heavy fermion superconductor UPd$_2$Al$_3$ has been investigated by Shimizu {\it et al.}~\cite{shimizu}
This is a good system to check our proposal.
The experimental data of the $\theta$ rotation of the specific heat $C/T$ at lower temperatures
exhibit the following characteristics in the oscillation patterns:
(1) In low fields the oscillation patterns of $C/T$ show a maximum at $\theta=0^{\circ }$
and monotonically decreases towards $\theta=90^{\circ}$.
(2) The maximum of the oscillation patterns at $\theta=0^{\circ }$ gradually moves to higher angles with 
increasing field.
Thus there appear two local minima at $\theta=0^{\circ }$ and $\theta=90^{\circ}$.
(3) In further high fields this maximum reaches at $\theta=90^{\circ}$.
The oscillation pattern becomes a monotonic curve whose minimum (maximum)
situated  at $\theta=0^{\circ }$ ($\theta=90^{\circ}$). Thus the oscillation pattern is reversed
between lower field and higher field ones.

Those three characteristics observed in UPd$_2$Al$_3$ are all reproduced by the 
present calculations for the horizontal linear line node case shown in Fig. 2(c).
Thus we can conclude that the nodal structure of UPd$_2$Al$_3$ has a linear 
line node at least by the following reasons:
(I) According to Shimizu {\it et al.},~\cite{shimizu} the azimuthal angle $\phi$ rotation of $C/T$
exhibits no detectable oscillation, indicating that there is no vertical line node.
(II) Since the local minima or flat bottom near $\theta=90^{\circ}$ is not observed,
 horizontal node must be a linear line node, rather than a quadratic one.
(III) As for the possible point nodes at the poles, it is difficult to discern the local minima at 
$\theta=0^{\circ }$ in view of the accuracy of the experimental data. Thus we cannot say whether  
there exists points nodes at the poles or not.

From these arguments, we may conclude that the chiral $d$-wave pairing $(k_x+ik_y)k_z$
is most likely.
It is fortunate that the whole oscillation  patterns predicted by the present calculations are
realized in UPd$_2$Al$_3$. One possible reason is the small $B_{\rm c2}$ anisotropy $B^c_{\rm c2}/B^{ab}_{\rm c2}\sim 1$
in this system.

\subsection{URu$_2$Si$_2$}

Kittaka {\it et al.}~\cite{kittaka} have performed the $\theta$ rotation of the specific heat experiment on 
a heavy fermion superconductor URu$_2$Si$_2$.
In low fields the oscillation patterns exhibit a maximum at $\theta=0^{\circ }$ and a
minimum at $\theta=90^{\circ}$.
As the field increases, a shoulder or kink structure appears at $\theta\sim45^{\circ}$,
reminiscent of the local maximum associated with the horizontal line node as seen from Fig. 2(c).
Upon further increasing field, the oscillation pattern becomes  a simple monotonic one
with the maximum at $\theta=0^{\circ }$. This simple pattern persists up to $B^c_{\rm c2}$.

The absence of the oscillation pattern reversal is due to the strong 
$B_{\rm c2}$ anisotropy where $B^c_{\rm c2}$ is suppressed by the Pauli paramagnetic effect.
This resembles the case explained in Sec.~III.B.1 where the crossing field disappears 
because the elongated Fermi surface makes the crossing field move to higher field.
Nevertheless, it is remarkable to observe the shoulder structure, which comes from the
nodal gap structure associated with a horizontal line node.
Since there are no local minima or flat bottom near $\theta=90^{\circ }$, the quadratic line node is unlikely.
Thus the chiral $d$-wave $(k_x+ik_y)k_z$ is the most probable pairing state.

\subsection{Cu$_x$Bi$_2$Se$_3$}

Recently the angle resolved specific heat experiment on a possible topological superconductor 
Cu$_x$Bi$_2$Se$_3$ has been reported and concluded that the two point nodes or gap minima exist
on the $k_x$-direction.~\cite{yonezawa}
By the exchange between the $k_z$- and $k_x$-directions, this situation is identical to our point node model 
shown in Fig.~6.
As mentioned, if the point nodes exist on the Fermi sphere, the oscillation patterns must be 
reversed as the field varies.
In addition, the local minima may appears at $\theta=0^{\circ }$ for the quadratic point node case.
Neither those features were observed by Yonezawa {\it et al.},~\cite{yonezawa}
where the same oscillation patterns persists for all fields up to $B^{ab}_{\rm c2}$ in low temperatures.
Thus although the observation of the two-fold oscillation pattern for the basal plane in the
hexagonal crystal is remarkable whose origin remains unknown,
we cannot support their conclusion that this two-fold oscillation is due to the point nodes or gap minima
from the present calculations. Since the oscillation patterns are independent of the field value,
it simply comes from the $B^{ab}_{\rm c2}$ anisotropy within the basal plane probed.

\subsection{UPt$_3$}

On the multi-phased heavy fermion superconductor UPt$_3$, the field angle resolved thermal conductivity measurement has been performed.~\cite{machida}
In addition to the remarkable two-fold oscillation patterns for the basal plane in the C-phase, the minima of the thermal conductivity are clearly observed at $\theta\sim 20^{\circ}$ and $160^{\circ}$ in the B- and C-phases [see Fig.~3 in Ref.~\onlinecite{machida}].
This may be an indication of the quadratic point nodes~\cite{sauls,nishira} or the off-symmetric line nodes.~\cite{tsutsumiUPt3}

\section{Summary}

We have proposed a spectroscopic method to probe the nodal gap structure and to pinpoint a nodal position in
momentum space by measuring the polar angle resolved ZDOS $N(\theta)$ via specific heat or thermal conductivity at low $T$.
The method is based on detection of the zero energy quasi-particles accumulated in the gap nodes in momentum space or equivalently
bounded in the vortex core in real space.
The ZDOS $N(\theta)$ exhibits the following features in a field range where the oscillation pattern is reversed:
(A) ``MAX''  structure, the global maxima in $N(\theta)$ drives this reversal, and (B) ``MIN'' structure, 
the local minima appear just near the angle of the nodal position in momentum space.
Those features are analyzed by quasiclassical Eilenberger 
theory beyond a simple Doppler shift picture and are explained clearly in terms of the
microscopic quasi-particles with zero energy in momentum space.

Thus, polar angle resolved measurement provides information on the nodal structure in typical tetragonal or hexagonal superconductors in addition to temperature and field dependence of specific heat or thermal conductivity.
Unless the oscillation reversal of the ZDOS is wiped off by the Fermi surface anisotropy, (A) the MAX structure and (B) the MIN structure or the flat-bottomed ZDOS around the node will be observed.
It is our hope that this spectroscopic method helps determining the pairing symmetry through
the nodal gap structure in various unconventional superconductors and pinpointing the momentum space position of Dirac node
and Weyl node in newly coming topological superconductors.

\section*{Acknowledgments}
We thank T.~Sakakibara, K.~Izawa, S.~Kittaka, Y.~Shimizu and Y.~Machida for informative discussions
on their experiments, which motivate this study.
A part of the numerical calculations was performed by using the HOKUSAI GreatWave supercomputer system in RIKEN.
Y.T. acknowledges financial support from the Japan Society for the Promotion of Science (JSPS).
This work was supported by KAKENHI Grant Nos. 15K17715, 15J05698, 15J01476, 15H05745, 15H02014, 26400360, and 25103716 from JSPS.

%\bibliography{aaa.bib}

\begin{thebibliography}{99}


\bibitem{volovik1}
G. E. Volovik, Superconductivity with lines of gap nodes: density of states in the vortex, JETP Lett. {\bf 58},  469 (1993).

\bibitem{volovik2}
G. E. Volovik, {\it The Universe in a Helium Droplet}
(Clarendon, Oxford, 2003).

\bibitem{matsuda}
Y. Matsuda, K. Izawa, and I. Vekhter, Nodal structure of unconventional superconductors probed
by the angle resolved thermal transport measurements, J. Phys.: Condens. Matter {\bf 18}, R705 (2006).

\bibitem{sakakibara}
T. Sakakibara, A. Yamada, J. Custers, K. Yano, T. Tayama, H. Aoki, and K. Machida,
Nodal structures of heavy Fermion superconductors probed by the specific-heat measurements in magnetic fields,
J. Phys. Soc. Jpn. {\bf 76}, 051004 (2007).

%\bibitem{sigrist}
%M. Sigrist and K. Ueda, 
%Phenomenological theory of unconventional superconductivity,
%Rev. Mod. Phys. {\bf 63}, 239 (1991).

\bibitem{vekhter}
I. Vekhter, P. J. Hirschfeld, J. P. Carbotte, and E. J. Nicol,
Anisotropic thermodynamics of d-wave superconductors in the vortex state,
Phys. Rev. B {\bf 59}, R9023 (1999).

\bibitem{kubert:1998}
C. K\"{u}bert and P. J. Hirschfeld,
Vortex contribution to specific heat of dirty d-wave superconductors: Breakdown of scaling,
Solid State Commun. {\bf 105}, 459 (1998).

\bibitem{nagai:2008b}
Y. Nagai and N. Hayashi,
Kramer--Pesch Approximation for Analyzing Field-Angle-Resolved Measurements Made in Unconventional Superconductors: A Calculation of the Zero-Energy Density of States,
Phys. Rev. Lett. {\bf 101}, 097001 (2008).

\bibitem{miranovic1}
P. Miranovi\'{c}, N. Nakai, M. Ichioka, and K. Machida, 
Orientational field dependence of low-lying excitations in the mixed state of unconventional superconductors,
Phys. Rev. B {\bf 68}, 052501 (2003).

\bibitem{vorontsov:2006}
A. Vorontsov and I. Vekhter,
Nodal Structure of Quasi-Two-Dimensional Superconductors Probed by a Magnetic Field,
Phys. Rev. Lett. {\bf 96}, 237001 (2006).

\bibitem{vorontsov:2007b}
A. Vorontsov and I. Vekhter,
Unconventional superconductors under a rotating magnetic field. I. Density of states and specific heat,
Phys. Rev. B {\bf 75}, 224501 (2007).

\bibitem{vorontsov:2007c}
A. Vorontsov and I. Vekhter,
Unconventional superconductors under a rotating magnetic field. II. Thermal transport,
Phys. Rev. B {\bf 75}, 224502 (2007).

\bibitem{miranovic2}
P. Miranovi\'{c},  M. Ichioka, K. Machida,  and N. Nakai, 
Theory of gap-node detection by angle-resolved specific heat measurement,
J. Phys.: Condens. Matter {\bf 17}, 7971 (2005).

\bibitem{hiragi}
M. Hiragi, K. M. Suzuki, M. Ichioka, and K. Machida, 
Vortex state and field-angle resolved specific heat oscillation for
$B\parallel ab$ in d-wave superconductors,
J. Phys. Soc. Jpn. {\bf 79}, 094709 (2010).

\bibitem{an}
K. An, T. Sakakibara, R. Settai, Y. Onuki, M. Hiragi, M. Ichioka, and K. Machida,
Sign reversal of field-angle resolved heat capacity oscillations in a heavy Fermion superconductor 
CeCoIn$_5$ and d$_{x^2-y^2}$ pairing symmetry,
Phys. Rev. Lett.  {\bf 104}, 037002 (2010).

\bibitem{nagai:2007}
Y. Nagai,Y. Kato, N. Hayashi, K. Yamauchi, and H. Harima
Calculated positions of point nodes in the gap structure of the borocarbide superconductor $\mathrm{Y}{\mathrm{Ni}}_{2}{\mathrm{B}}_{2}\mathrm{C}$,
Phys. Rev. B {\bf 76}, 214514 (2007).

\bibitem{nagai:2009}
Y. Nagai,Y. Kato, N. Hayashi, K. Yamauchi, and H. Harima
Field angle dependence of the zero-energy density of states in unconventional superconductors: analysis of the borocarbide superconductor YNi$_2$B$_2$C,
J. Phys.: Conf. Series {\bf 150}, 052177 (2009).

\bibitem{nakatsuji} 
See for example, T. Kondo, et al, Quadratic Fermi node in a 3D strongly correlated semimetal, Nature Commun.
 {\bf 6}, 10042 (2015).


\bibitem{fu}
L. Fu and E. Berg, 
Odd-parity topological superconductors: theory and application to Cu$_x$Bi$_2$Se$_3$,
Phys. Rev. Lett. {\bf 105}, 097001 (2010).

\bibitem{fan}
Shengyuan A. Yang, Hui Pan, and Fan Zhang, 
Dirac and Weyl superconductors in three dimensions,
Phys. Rev. Lett. {\bf 113}, 046401 (2014).


\bibitem{xu}
Yong Xu, Fan Zhang, and Chuanwei Zhang, 
Structured Weyl points in Fulde-Ferrell superfluids,
Phys. Rev. Lett. {\bf 115}, 265304 (2015).

\bibitem{haldane}
Yi Li and F. D. M. Haldane, Topological nodal Cooper pairing in doped Weyl metals,
arXiv:1510.01730.

\bibitem{kittaka}
S. Kittaka, Y. Shimizu, T. Sakakibara, Y. Haga, E. Yamamoto,
Y. Onuki, Y. Tsutsumi, T. Nomoto, H. Ikeda, and K. Machida, 
Evidence for chiral d-wave superconductivity in URu$_2$Si$_2$ from the field-angle variation of its specific heat,
J. Phys. Soc. Jpn. {\bf 85}, 033704 (2016).

\bibitem{shimizu}
Y. Shimizu, S. Kittaka, T. Sakakibara, Y. Tsutsumi, T. Nomoto, H. Ikeda, K. Machida,
Y. Homma, and D. Aoki, Omni-directional measurements of angle-resolved heat capacity for complete
detection of superconducting gap structure in the heavy-fermion antiferromagnet UPd$_2$Al$_3$,
Phys. Rev. Lett. {\bf 117}, 037001 (2016).

\bibitem{eilenberger}
G. Eilenberger,
Determination of $\kappa_1(T)$ and $\kappa_2(T)$ for type-II superconductors with arbitrary impurity concentration,
Phys. Rev. {\bf 153}, 584 (1967).

\bibitem{ichioka1}
M. Ichioka, N. Hayashi, and K. Machida, 
Local density of states in the vortex lattice in a type-II superconductor,
Phys. Rev. B {\bf 55}, 6565 (1997).

\bibitem{ichioka2}
M. Ichioka, A. Hasegawa, and K. Machida, 
Vortex lattice effects on low-energy excitations in d-wave and s-wave superconductors,
Phys. Rev. B {\bf 59}, 184 (1999).

\bibitem{ichioka3}
M. Ichioka, A. Hasegawa, and K. Machida, 
Field dependence of the vortex structure in d-wave and s-wave superconductors,
Phys. Rev. B {\bf 59}, 8902 (1999).

\bibitem{tsutsumiUPt3}
Y. Tsutsumi, K. Machida, T. Ohmi, and M. Ozaki,
A Spin Triplet Superconductor UPt$_3$,
J. Phys. Soc. Jpn. {\bf 81}, 074717 (2012).

\bibitem{Nagai2006}
Y. Nagai, Y. Ueno, Y. Kato, and N. Hayashi,
Analytical formulation of the local density of states around a vortex core in unconventional superconductors,
J. Phys. Soc. Jpn. {\bf 75}, 104701 (2006).

\bibitem{Nagai2011}
Y. Nagai, H. Nakamura, and M. Machida,
Superconducting gap function in the organic superconductor (TMTSF)$_2$ClO$_4$ with anion ordering; 
First-principles calculations and quasiclassical analysis for angle-resolved heat capacity,
Phys. Rev. B {\bf 83}, 104523 (2011).

\bibitem{nakai}
N. Nakai,  P. Miranovi\'{c}, M. Ichioka, and K. Machida, 
Field dependence of the zero-energy density of states around vortices in an anisotropic-gap superconductor,
Phys. Rev. B {\bf 70}, 100503(R) (2004).

\bibitem{phasespace}
The low energy excitation spectra are 
given by $N(E)\propto E^2$ for linear point nodes, $N(E)\propto E$ for a linear line node and quadratic point nodes,
and $N(E)\propto \sqrt{|E|}$ for a quadratic line node.

\bibitem{mishra:2011}
V. Mishra, S. Graser, and P. J. Hirschfeld,
Transport properties of three-dimensional extended $s$-wave states in Fe-based superconductors,
Phys. Rev. B {\bf 84}, 014524 (2011).

\bibitem{yonezawa}
S. Yonezawa, K. Tajiri, S. Nakata, Y. Nagai, Z. Wang, Y. Ando, and Y. Maeno, 
Thermodynamic evidence for nematic superconductivity in Cu$_{\rm x}$Bi$_{2}$Se$_{3}$,
arXiv:1602.08941.

\bibitem{machida}
Y. Machida, A. Itoh, Y. So, K. Izawa, Y. Haga, E. Yamamoto, N. Kimura, Y. Onuki, Y. Tsutsumi, and K. Machida,
Twofold spontaneous symmetry breaking in the heavy-Fermion superconductor UPt$_3$,
Phys. Rev. Lett.  {\bf 108}, 157002 (2012).

\bibitem{sauls}
M. J. Graf, S. -K. Yip, and J. A. Sauls, 
Identification of the orbital pairing symmetry in UPt$_3$,
Phys. Rev. B {\bf 62}, 14393 (2000).

\bibitem{nishira}
K. Machida, T. Nishira, and T. Ohmi, 
Orbital symmetry of a triplet pairing in a heavy Fermion superconductor UPt$_3$,
J. Phys. Soc. Jpn. {\bf 68}, 3364 (1999).
%Y. Tsutsumi, M. Ishikawa, T. Kawakami, T. Mizushima, M. Sato, M. Ichioka, and K. Machida,
%UPt$_3$ as a topological crystalline superconductor,
%J. Phys. Soc. Jpn. {\bf 82}, 113707 (2013).

\end{thebibliography}
%\end{document}
%%%%%%%%%%%%%%%%%%%%%%%%%%%%%%%%%%%%%%%%%%%

\end{document}